\documentclass[a4paper,12pt]{article}
\usepackage{amsmath}
\usepackage{epsfig}

\begin{document}

\begin{titlepage}
\begin{center}
\hfill    CERN-PH-TH/2005-018\\
~{} \hfill hep-ph/0502yyy\\
\vskip 1cm

{\large \bf   CP-odd invariants in models with \\ 
several Higgs doublets}

\vskip 1cm

Gustavo C. Branco\footnote{gustavo.branco@cern.ch and 
gbranco@cftp.ist.utl.pt}  \ $^{4}$
M. N. Rebelo
\footnote{margarida.rebelo@cern.ch and
rebelo@ist.utl.pt} \ $^4 $
and 
J. I. Silva-Marcos\footnote{Joaquim.Silva-Marcos@cern.ch} \ 
\footnote{On leave of absence from 
Departamento de F\'\i sica and Centro de F\' \i sica Te\' orica
de Part\' \i culas (CFTP),
Instituto Superior T\' ecnico, Av. Rovisco Pais, P-1049-001 Lisboa,
Portugal.}
\vskip 0.05in

{\em Physics Department, Theory, CERN, CH-1211 Geneva 23, 
Switzerland.}
\end{center}

\vskip 3cm

\begin{abstract}
We present CP-odd Higgs-basis invariants, which can be used
to signal CP violation in a multi-Higgs system, written
in an arbitrary Higgs basis. It is shown through specific examples
how these CP-odd invariants can also be useful to determine
the character of CP breaking (i.e. whether it is hard or soft
CP breaking) in a given Higgs Lagrangian. We analyse in detail
the cases of two and three Higgs doublets.
\end{abstract}

\end{titlepage}

\section{Introduction}
One of the most fundamental open questions in particle physics
is discovering what the mechanism responsible for gauge-symmetry 
breaking is. In the Standard Model (SM), the breaking 
of the $SU(2)\times U(1)\times
SU(3)_{c}$ symmetry is achieved by the Higgs mechanism, 
realized through the
introduction of one Higgs doublet. However, there is no fundamental
reason for having only one Higgs doublet, and multi-Higgs
systems are required by a large class of models, including 
supersymmetric models as well as models where CP is spontaneously
broken \cite{Lee:1973iz} \cite{Branco}. In the SM with only one Higgs
doublet, the Higgs sector does not contain new sources of CP violation,
but this is no longer true in models with more than one Higgs doublet.
This has prompted several recent analyses on the impact of
multi-Higgs in the physics at future colliders \cite{colliders}. 

In this paper, we study CP violation in multi-Higgs extensions of the
SM, addressing in particular the question of finding the conditions for a 
given Higgs potential to violate CP at the Lagrangian level.
This question is not trivial, since in a model with
more than one Higgs doublet, one has the freedom to make
Higgs-basis transformations that do not change the physical 
content of the model, but do change both the quadratic and the quartic
couplings. Couplings that are complex in one Higgs basis
may become real in another basis. A similar difficulty arises
when one investigates whether, in a given model, there is
soft or hard CP breaking. Indeed a given model may have complex quartic 
couplings in one Higgs basis, while they may all become real 
in another basis, with only the quadratic couplings 
now complex, thus indicating that in that particular model CP is only softly 
broken.

The above intrinsic ambiguities motivate the study of conditions for CP
invariance expressed in terms of CP-odd Higgs-basis invariants. We 
analyse in detail the cases of two and three Higgs doublets and
briefly describe the case of an arbitrary number of Higgs doublets.

\section{The Higgs Lagrangian}

Let us consider the Standard $SU(2)\times U(1)\times
SU(3)_{c}$ Model with $n_{d}$ Higgs doublets. For the moment we 
consider only the Higgs sector. The most general renormalizable polynomial
consistent with the gauge invariance is given by:
\begin{equation}
{\mathcal{L}}_{\phi }=Y_{ab}\ \phi _{a}^{\dagger }\phi _{b}+Z_{abcd}\ 
\left( \phi _{a}^{\dagger }\phi _{b}\right) \left( \phi _{c}^{\dagger }\phi
_{d}\right),  \label{laghiggs}
\end{equation}
where repeated indices are summed. Hermiticity of 
${\mathcal{L}}_{\phi }$ implies:
\begin{equation}
\begin{array}{ccc}
Y_{ab}^{\ast }=Y_{ba} & \quad ;\quad & Z_{abcd}^{\ast }=Z_{badc} \ .
\end{array}
\label{hermzy}
\end{equation}
Furthermore, by construction it is obvious that:
\begin{equation}
Z_{abcd}=Z_{cdab} \ .  \label{symz}
\end{equation}
It is important to note that one can make a Higgs-basis
transformation (HBT) defined by:
\begin{equation}
\begin{array}{ccc}
\phi _{a}\overset{\mathrm{HBT}}{\longrightarrow }
\phi _{a}^{\prime }=V_{ai}\ \phi _{i}
& \quad ,\quad & \phi _{a}^{\dagger }
\overset{\mathrm{HBT}}{\longrightarrow }\left(
\phi ^{\prime }\right) _{a}^{\dagger }=V_{ai}^{\ast} \ 
\left( \phi ^{\prime }\right)
_{i}^{\dagger }\ ,
\end{array}
\label{trafowb}
\end{equation}
where $V$ is an $n_{d}\times n_{d}$ 
unitary matrix acting in the space of
Higgs doublets. Under a HBT, the physics content of ${\mathcal{L}}_{\phi }$
does not change, but the couplings $Y$ and $Z$ transform as:
\begin{equation}
\begin{array}{c}
Y_{ab}\ \overset{\mathrm{HBT}}{\longrightarrow }
\ Y_{ab}^{\prime }\ =V_{am}\ Y_{mn}\
V^{\dagger }_{nb}\  \\ 
\\ 
Z_{abcd}\overset{\mathrm{HBT}}{\longrightarrow }
\ Z_{abcd}^{\prime }=V_{am}\ V_{cp}\
Z_{mnpq}\ \ V^{\dagger }_{nb}\ V^{\dagger }_{qd}.\ 
\end{array}
\label{wbyz2}
\end{equation}
We now address the following question: What are the necessary and
sufficient conditions for ${\mathcal{L}}_\phi $ to be CP-invariant?
Note that an entirely analogous question has been asked in the
framework of the SM, for the Yukawa sector, where it has been shown
\cite{Bernabeu:1986fc} that a necessary condition for CP invariance
at the Lagrangian level in the SM is given by:
\begin{equation}
\mathrm{Tr} \ [\ g_u g_u^\dagger, \ g_d g_d^\dagger \ ] ^3 =0,
\label{yuk}
\end{equation}
where $g_u$, $g_d$ denote the Yukawa couplings involving
$u_R$ and $d_R$ respectively; after gauge symmetry breaking, these
lead to the  quark masses for the up and down quarks.
The condition of Eq.~(\ref{yuk}) is a necessary condition
for CP invariance for an arbitrary number of generations
and, for three generations, is also a sufficient condition for
CP invariance.

In order to answer the above question for a multi-Higgs sector, 
we follow the general method 
proposed in \cite{Bernabeu:1986fc}, 
applied to specific models in \cite{many1} 
 and described in detail in \cite%
{Branco:1999fs}. The simplest approach consists in considering the most
general CP transformation of the Higgs fields, which leaves the
Higgs kinetic energy term invariant,
and investigate what restrictions are implied for
the couplings $Y$ and $Z$, by CP invariance. The most general CP
transformation that leaves the kinetic energy invariant is:
\begin{equation}
\begin{array}{ccc}
\phi _{a}\overset{\mathrm{CP}}{\longrightarrow }
U_{ai}\ \phi _{i}^{\ast } & \quad
;\quad & \phi _{a}^{\dagger }\overset{\mathrm{CP}}{\longrightarrow }
U_{ai}^{\ast }\
\phi _{i}^{T}
\end{array}
\label{trafocp}
\end{equation}
where $U$ is an $n_{d}\times n_{d}$ unitary matrix operating in Higgs
doublets space. It can be readily seen, using Eqs.~(\ref{laghiggs}) 
and (\ref{trafocp}), that the necessary and sufficient condition for 
${\mathcal{L}}_{\phi }$ to conserve CP is that the following 
relations be satisfied:
\begin{equation}
\begin{array}{c}
\left( Y^{\ast }\right) _{ab}=\ U_{am}^{\dagger }\ Y_{mn}\ U_{nb}\  \\ 
\\ 
\left( Z^{\ast }\right) _{abcd}=\ \ U_{am}^{\dagger }\ U_{cp}^{\dagger }\ \
Z_{mnpq}\ U_{nb}\ U_{qd}.
\end{array}
\label{cpzy}
\end{equation}
In other words, for a given ${\mathcal{L}}_{\phi }$, written as in
Eq.~(\ref{laghiggs}), the necessary and sufficient 
condition for ${\mathcal{L}}_{\phi }$ to be CP-invariant is the existence 
of an $n_{d}$ dimensional
unitary matrix $U$, 
satisfying Eqs.~(\ref{cpzy}). Obviously, the above criterion
is HBT-invariant. In order to show that this is indeed the case, let us
assume that a solution $U$ exists for ${\mathcal{L}}_{\phi }$, written in a
certain basis. Then a solution $U^{\prime }$ will also exist in another
basis related to the initial one through Eq.~(\ref{trafowb}). 
From Eqs.~(\ref{trafowb}), (\ref{wbyz2}) and (\ref{cpzy}) 
it follows that, in the primed
basis, the solution is given by:
\begin{equation}
U^{\prime }=V\ U\ V^{T}  \label{uprime}.
\end{equation}
Although Eqs.~(\ref{cpzy}) provide necessary and sufficient
conditions for CP invariance in the Higgs sector, valid for an arbitrary
number of Higgs doublets, they are not very practical, since they require
the search for a solution for the unitary matrix $U$. It is thus useful to
consider quantities which, on the one hand are HBT-invariant and, on the
other hand, are constrained to vanish by CP invariance. In the case
of the Higgs sector, an analysis \cite{higgs}
has been done in terms of HBT
invariants which involve vacuum expectation values (vev's) of Higgs fields.
These HBT invariants are very useful at low energies in the analysis of
various CP-violating phenomena, but they do not answer directly the question
we address in this paper, to wit, what the HBT-invariant conditions for
CP invariance of ${\mathcal{L}}_{\phi }$, prior to gauge symmetry breaking are.
The relevance of this question has been emphasized in \cite{cern}.
Furthermore, at high energies,  $SU(2)\times U(1)\times SU(3)_{c}$ is
not broken, and therefore HBT
invariants involving vev's are no longer relevant. The CP-odd HBT
invariants, which we consider in this paper, are expressed directly in terms
of the couplings $Y$ and $Z$ of the Lagrangian and thus continue to be
relevant at high energies. Note that CP-odd HBT invariants that do not
vanish at high energies are specially useful in the analysis of
baryogenesis, including baryogenesis through leptogenesis.

\section{Construction of CP-odd Higgs-basis \\ transformation invariants}

In the construction of HBT invariants, it is useful to adopt a more compact
notation, which renders more transparent the invariance of various quantities
under HBT. We define:%
\begin{equation}
\begin{array}{c}
Y_{a}^{b} \equiv Y_{ab} \\ 
\\ 
Z_{ac}^{bd} \equiv Z_{abcd} \ ,
\end{array}
\label{uplow}
\end{equation}
where the upper indices are those that transform with $V^{\dagger }$
under a HBT, as specified in Eqs.~(\ref{wbyz2}).

Next we consider separately the cases of two and three Higgs doublets.

\subsection{Two Higgs doublets}

\subsubsection{The general case}

For later convenience, and in order to settle the notation, we write
explicitly the most general Higgs potential for two Higgs doublets as:%
\begin{equation}
\begin{array}{l}
V_{H_{2}}=m_{1}\ \phi _{1}^{\dagger }\phi _{1}+p\ e^{i\varphi }\ \phi
_{1}^{\dagger }\phi _{2}+p\ e^{-i\varphi }\ \ \phi _{2}^{\dagger }\phi
_{1}+m_{2}\ \phi _{2}^{\dagger }\phi _{2}+ \\ 
\\ 
+a_{1}\ \left( \phi _{1}^{\dagger }\phi _{1}\right) ^{2}+a_{2}\ \left( \phi
_{2}^{\dagger }\phi _{2}\right) ^{2}+b\ \left( \phi _{1}^{\dagger }\phi
_{1}\right) \left( \phi _{2}^{\dagger }\phi _{2}\right) +b^{\prime }\ \left(
\phi _{1}^{\dagger }\phi _{2}\right) \left( \phi _{2}^{\dagger }\phi
_{1}\right) + \\ 
\\ 
+c_{1}\ e^{i\theta _{1}}\ \left( \phi _{1}^{\dagger }\phi _{1}\right) \left(
\phi _{2}^{\dagger }\phi _{1}\right) +c_{1}\ e^{-i\theta _{1}}\ \left( \phi
_{1}^{\dagger }\phi _{1}\right) \left( \phi _{1}^{\dagger }\phi _{2}\right) +
\\ 
\\ 
+c_{2}\ e^{i\theta _{2}}\ \left( \phi _{2}^{\dagger }\phi _{2}\right) \left(
\phi _{2}^{\dagger }\phi _{1}\right) +c_{2}\ e^{-i\theta _{2}}\ \left( \phi
_{2}^{\dagger }\phi _{2}\right) \left( \phi _{1}^{\dagger }\phi _{2}\right) +
\\ 
\\ 
+d\ e^{i\delta }\ \left( \phi _{1}^{\dagger }\phi _{2}\right) ^{2}+d\
e^{-i\delta }\ \left( \phi _{2}^{\dagger }\phi _{1}\right) ^{2}.
\end{array}
\label{higgs2}
\end{equation}
We have written the most general Higgs potential for two Higgs doublets,
displaying the phase dependence explicitly. It is clear that this 
potential contains an excess of parameters. This is entirely
analogous to the situation encountered in the fermion
sector where, for example, the identification of the correct number
of independent parameters contained in Yukawa couplings requires
that one go to the weak basis, where one of the Yukawa coupling matrices 
is diagonal and real, while the other one is Hermitian, with only one
rephasing-invariant phase. Similarly in the Higgs sector a basis 
can be choosen, without loss of generality, where the quadratic terms 
are diagonal and furthermore, of the three remaining phases 
$\theta_1$, $\theta_2$, $\delta$, one can still eliminate
one by rephasing one of the Higgs fields. This
means that in a two Higgs doublets system there are only 
two independent CP-violating phases.

We want to find the necessary and sufficient conditions for CP
invariance in terms of CP-odd HBT invariants. From the above 
counting, we expect to find two such conditions. 
As previously mentioned, in
Ref.~\cite{higgs} CP-odd invariants were constructed involving $v_{a}\equiv
\left\langle 0\left\vert \phi _{a}\right\vert 0\right\rangle $. The simplest
invariants of this type are:
\begin{equation}
\begin{array}{l}
\tilde{I_{1}}\equiv Y_{a}^{b}\;Z_{bc}^{cd}\;W_{d}^{a} \\ 
\\ 
\tilde{I_{2}}\equiv W_{a}^{b}\;Y_{b}^{c}\;W_{d}^{e}\;Y_{e}^{f}\;Z_{cf}^{ad} \ ,
\end{array}
\label{invvev}
\end{equation}
where $W_{a}^{b}\equiv W_{ab}\equiv v_{a}v^{\ast }_{b}$. CP invariance 
requires Im($\tilde{I_{i}}$) to vanish. 
One could be tempted to consider HBT invariants
analogous to $\tilde{I_{1}}$ and $\tilde{I_{2}}$, simply replacing $%
W_{i}^{j} $ by $Y_{i}^{j}$. However, it can be readily verified that these
new HBT invariants are automatically real. In fact, with this replacement
both $\tilde{I_{i}}$ become traces of products of two Hermitian matrices and
such traces are always real, independently of the number of Higgs doublets. 
\newline
\textbf{A set of necessary and sufficient conditions for CP invariance}:
Next we show that in the case of two Higgs doublets, necessary 
and sufficient conditions 
for  ${\mathcal{L}}_{\phi } \equiv  V_{H_{2}}$
to be CP-invariant are:
\begin{equation}
\begin{array}{l}
I_{1}\equiv \mathrm{Tr}[Y\ Z_{Y}\ \widehat{Z}-\widehat{Z}\ Z_{Y}\ Y]=0 \\ 
\\ 
I_{2}\equiv \mathrm{Tr}[Y\ Z_{2}\ \widetilde{Z}-\widetilde{Z}\ Z_{2}\ Y]=0 \ ,
\end{array}
\label{inv}
\end{equation}
where we have defined the following $n_{d}\times n_{d}$ 
Hermitian matrices{\footnote{
For this particular case, $n_{d}=2$}.}: 
\begin{equation}
\begin{array}{ccc}
\widehat{Z}_{a}^{b}\equiv Z_{am}^{bm} &  & \widetilde{Z}_{a}^{b}\equiv
Z_{am}^{mb} \\ 
&  &  \\ 
\left( Z_{Y}\right) _{a}^{b}\equiv Z_{an}^{bm}Y_{m}^{n} &  & \left(
Z_{2}\right) _{a}^{b}\equiv Z_{an}^{pm}Z_{mp}^{nb}
\end{array}
\label{contrzy}
\end{equation}
The conditions of Eqs.~(\ref{inv}) are a chosen set of equations,
which can be derived from Eq.~(\ref{cpzy}), by using the property of
invariance of the trace under similarity transformations. 
Therefore, they are necessary conditions for CP invariance
for an arbitrary number of Higgs doublets. 
In order to prove that they are also sufficient in the case of
two Higgs doublets, let us
consider the explicit form of the potential 
${\mathcal{L}}_{\phi }=V_{H_{2}}$,
written above for the case of two Higgs doublets, where all phases allowed by
hermiticity are written explicitly. 
Since the conditions of  Eq.~(\ref{inv}) are written in 
terms of HBT invariants, they can be computed in any appropriately
chosen HB. In the HB basis, where the quadratic terms are diagonal, these
conditions can be written as:
\begin{equation}
\begin{array}{l}
I_{1} \equiv \frac{i}{2}c_{1}c_{2}(m_{1}-m_{2})^{2}\sin 
(\theta _{2}-\theta _{1}) = 0
\\ 
\\ 
I_{2} \equiv \frac{i}{2}\ (m_{1}-m_{2})\ [\ dc_{1}^{2}\sin 
(\delta +2\theta
_{1})+dc_{2}^{2}\sin (\delta +2\theta _{2})+ \\ 
\\ 
+2dc_{1}c_{2}\sin (\delta +\theta _{1}+\theta
_{2})+c_{1}c_{2}(a_{2}-a_{1})\sin (\theta _{1}-\theta _{2})\ ] = 0 \ .
\end{array}
\label{higgs2i}
\end{equation}
Note that, for simplicity, we have omitted the primes in the coefficients,
which obviously assume new values after the HBT that diagonalizes
the quadratic terms.
As mentioned above, by making a phase redefinition of the
$\phi_i$, one of the three phases $\theta_1$,
$\theta_2$, $\delta$ can be eliminated. 
We choose $\delta = 0$. 
From the first of Eqs.~(\ref{higgs2i}) we obtain:
\begin{equation}
 \theta_2 = \theta_1 \qquad  \mbox{or} \qquad \theta_2 =  \theta_1 + \pi \ ,
\label{thetas}
\end{equation}
where we have assumed non-degeneracy of $m_i$. Taking for definiteness 
$\theta_2 = \theta_1$, the condition $I_2 = 0$, implies:
\begin{equation}
I_{2} \equiv \frac{i}{2}\ (m_{1}-m_{2})\ \ d (c_{1} + c_{2} )^{2}
\sin (2\theta_{1}) =0,
\label{simp}
\end{equation}
leading to two solutions:
\begin{equation}
\theta_1 = \theta_2 =0  \qquad  \mbox{or} \qquad \theta_1 = 
\theta_2 = \pi/2 \ .
\label{solu}
\end{equation}
The case $\theta_1 = \theta_2 =0 $ obviously corresponds to 
${\mathcal{L}}_{\phi }$ CP-invariant, with $\phi_1$, $\phi_2$
transforming trivially under CP, i.e. the matrix $U$ in 
Eq.~(\ref{trafocp}) is a $2 \times 2$ identity matrix.
It can be easily checked that the solution 
$\theta_1 = \theta_2 = \pi/2 $ also corresponds to
a CP-invariant ${\mathcal{L}}_{\phi }$ with the matrix $U$
given by:
\begin{equation}
U = \left(
\begin{array}{cc}
1 & 0 \\
0 & -1 
\end{array}
\right).
\label{uuuu}
\end{equation}
Therefore this solution corresponds to two Higgs doublets
with opposite CP parities.
In our proof, we have assumed that there is no degeneracy of $m_i$.
It is clear from Eqs.~(\ref{higgs2i}) that in the degenerate 
limit $m_1 = m_2$, both $I_1$, $I_2$ vanish and yet there is
the possibility of CP violation.
One can easily construct a HBT invariant that does not
trivially vanish in the degenerate limit; an example is:   
\begin{equation}
I_{3}=\mathrm{Tr}[Z_{2}\ Z_{3}\ \widehat{Z}-\widehat{Z}\ Z_{3}\ Z_{2}] \ ,
\label{invzz}
\end{equation}
where $Z_{3}$ is also an $n_{d}\times n_{d}$ Hermitian matrix given by:
\begin{equation}
\left( Z_{3}\right) _{a}^{b}=Z_{ar}^{mp}Z_{mn}^{rs}Z_{ps}^{nb} \ .  \label{z3}
\end{equation}
The explicit form of $I_{3}$ in terms of parameters of the Lagrangian is
given by:
\begin{equation}
\begin{array}{r}
I_{3}=A\sin (\theta _{1}-\theta _{2})+B\sin (2(\theta _{1}-\theta
_{2}))+C\sin (\delta +\theta _{1}+\theta _{2})+ \\ 
\\ 
+\underset{ij=1,2}{\sum }C_{ij}\sin (2\delta +\theta _{i}+3\theta
_{j})+D_{ij}\sin (\delta +3\theta _{i}-\theta _{j}) \ ,
\end{array}
\label{invzz1}
\end{equation}
where the $A$, $B$, $C$'s and $D$'s are polynomials of degree 6 
in $a$'s, $b$'s, $c$'s and $d$, the parameters of the quartic 
terms of the two Higgs
potential in Eq.~(\ref{higgs2}).

\subsubsection{Soft CP breaking with two Higgs doublets}
Let us now consider the soft CP-breaking case, 
in which CP is explicitly broken
and there is a Higgs basis where all quartic couplings of the potential are
real and the coefficient $Y_{12}\equiv Y_{1}^{2}=p\;e^{i\varphi }$ 
complex. This case may be interesting in the context of models with explicit
CP violation and suppressed Higgs flavour-neutral currents 
\cite{Branco:1985aq}. Next, we show that
the CP-odd HBT invariants that we have considered are also 
useful to find out whether, in a given model, there is hard 
or soft CP breaking. It is clear that if a CP-odd invariant, 
built exclusively from quartic couplings, does not vanish, 
then one has hard CP breaking. On the other hand, one may have 
complex $Z$ couplings in a given HB and yet CP be only softly broken.
In this case HBT invariants such as $I_3$ would vanish and 
only HBT invariants including combinations of $Y$'s and $Z$'s
could signal CP violation.
We illustrate such a situation through a simple
example. Let us consider the following Higgs potential:
\begin{eqnarray}
V^\prime_{H_{2}} & = & m_{1}\ \phi _{1}^{\dagger }\phi _{1}+
m_{2}\ \phi _{2}^{\dagger }\phi _{2} + \nonumber\\ 
& + & a \left[ \ \left( \phi _{1}^{\dagger }
\phi _{1}\right) ^{2}+ \ \left( \phi
_{2}^{\dagger }\phi _{2}\right) ^{2} \right]
 + b\ \left( \phi _{1}^{\dagger }\phi 
_{1}\right) \left( \phi _{2}^{\dagger }\phi _{2}\right) +
\label{hid} \\ 
& + & c \ \left[ \ e^{i\theta}\ \left( \phi _{1}^{\dagger }
\phi _{1}\right) \left(
\phi _{1}^{\dagger }\phi _{2}\right)  
 + \  e^{i\theta}\left( \phi
_{2}^{\dagger }\phi _{2}\right) \left( \phi _{2}^{\dagger }\phi _{1}\right) +
\mathrm{h.c.}\right] + \nonumber \\
& + &d \ \left[\ e^{i\delta }\ \left( \phi _{1}^{\dagger }
\phi _{2}\right) ^{2}+ \mathrm{h.c.} \right]. \nonumber
\end{eqnarray}
An explicit evaluation of $I_i$ ($i$ = 1,2,3) gives:
\begin{equation}
\begin{array}{l}
I_{1}\equiv \mathrm{Tr}[Y\ Z_{Y}\ \widehat{Z}-\widehat{Z}\ Z_{Y}\ Y]=0 \\ 
I_{2}\equiv \mathrm{Tr}[Y\ Z_{2}\ \widetilde{Z}-\widetilde{Z}\ Z_{2}\ Y]=
2 c^2 d \ (m_1 - m_2) \ \sin (\delta + 2 \theta) \\
I_3 \equiv \mathrm{Tr}[Z_{2}\ Z_{3}\ \widehat{Z}-
\widehat{Z}\ Z_{3}\ Z_{2}] = 0 \ .
\end{array}
\label{tres}
\end{equation}
The fact that $I_3 = 0$, while $I_2 \neq 0$, provides a hint that CP
is only softly broken. It can be verified that this is indeed the case,
because the quartic couplings, by themselves, conserve CP, provided one
chooses in Eq.~(\ref{trafocp}) the following $U$ matrix:
\begin{equation}
U = \left(
\begin{array}{cc}
0 & 1 \\
1 & 0 
\end{array}
\right) \ .
\label{hidd}
\end{equation}
One may say that ${\mathcal{L}}_{\phi } = V^\prime_{H_{2}}$ of
Eq.~(\ref{hid}) corresponds to ``hidden'' soft CP breaking.

\subsection{Three Higgs doublets}

\subsubsection{The general case}
Three Higgs doublets were considered in an attempt to introduce
CP violation in an extension of the SM with natural flavour
conservation (NFC) \cite{Glashow:1976nt} in the Higgs sector.
If one introduces NFC in a two Higgs doublet system, through an exact
symmetry of the Lagrangian under which each Higgs doublet
transforms differently, there is no longer the possibility of
CP violation in the Higgs sector. Is is clear from Eq.~(\ref{higgs2})
that such a symmetry, in the case of two Higgs doublets, would forbid all
terms in $c_{i}$ as well as the quadratic non-diagonal term, so that the
phase of the term in $d$ could then be rephased away. On the other 
hand, it was shown that with three Higgs doublets it is possible 
to violate CP in the Higgs sector either at the Lagrangian level
\cite{Weinberg:1976hu} or spontaneously \cite{Branco}, 
while having NFC.  The general three Higgs doublets system, 
without extra symmetries, is rather complicated, due to the large
number of parameters involved. It can be readily verified that
for $n_d$ Higgs doublets the total number of CP-violating phases
contained in the Higgs potential is:
\begin{equation}
N_{\mathrm{phases}} = \frac{1}{4} \left[ n_d^2 ( n_d^2 -1) \right] 
- (n_d -1).
\label{cont}
\end{equation}
The second term in Eq.~(\ref{cont}) results from the 
fact that one can eliminate $(n_d -1)$ phases by rephasing the Higgs
fields. From Eq.~(\ref{cont}) it follows that the number of independent 
CP-violating phases grows fast with increasing $n_d$. One has
0, 2, 16 independent phases for $n_d = 1, 2, 3$ respectively.
In view of this, we only point out some of the novel generic
features which arise in the case of three Higgs doublets and then
consider specific models. It is clear that the CP-odd invariants 
$I_1$, $I_2$ defined in Eqs.~(\ref{inv}) provide necessary 
conditions for CP invariance for an arbitrary number of generations. 
However, there are simpler CP-odd invariants, which are irrelevant
for the case of two Higgs doublets (since they trivially vanish
in that case) but are useful in the case of three Higgs doublets.
A simple CP-odd invariant of this class is:
\begin{equation}
I_{s}=\mathrm{Tr}\left( [Y\ ,\ \widetilde{Z}]^{3}\right) \ .  \label{higgs3i}
\end{equation}
An equivalent invariant can be obtained by replacing $\tilde{Z}$
by $\hat{Z}$ as defined in Eq.~(\ref{contrzy}). 
The vanishing of $I_{s}$ is
a non-trivial necessary condition for CP invariance in the case of 
three or more Higgs doublets. It is analogous to the condition
obtained in the quark sector, given in Eq.~(\ref{yuk}). In the HB 
where $Y$ is diagonal, one obtains:
\begin{equation}
I_{s} = 6i \ (Y_{22} -Y_{11})\ (Y_{33} -Y_{11}) \ (Y_{33} -Y_{22}) \
\mathrm{Im}(\widetilde{Z}_{12}\widetilde{Z}_{23}\widetilde{Z}_{31}) \ .
\label{trc}
\end{equation}

\subsubsection{Weinberg three Higgs doublets model}
In the three Higgs doublets model proposed by 
Weinberg \cite{Weinberg:1976hu}, a
$Z_{2}\times Z_{2}\times Z_{2}$
symmetry, under separate reflections of the Higgs doublets of the form $\phi
_{i}\rightarrow -\phi _{i}$, together with an appropriately chosen
transformation for the quark fields, ensures NFC and leads 
to a strong reduction in the number of parameters. The Higgs
potential is given by:
\begin{equation}
\begin{array}{r}
V=\underset{i=123}{\sum }m_{i}\ \phi _{i}^{\dagger }\phi _{i}+a_{ii}\ \left(
\phi _{i}^{\dagger }\phi _{i}\right) ^{2}+\underset{i<j}{\sum }2b_{ij}\
\left( \phi _{i}^{\dagger }\phi _{i}\right) \ \left( \phi _{j}^{\dagger
}\phi _{j}\right) + \\ 
\\ 
+2c_{ij}\ \left( \phi _{i}^{\dagger }\phi _{j}\right) \ \left( \phi
_{j}^{\dagger }\phi _{i}\right) +\left[ d_{ij}\ e^{i\theta _{ij}}\ \left(
\phi _{i}^{\dagger }\phi _{j}\right) ^{2}+h.c.\right].
\end{array}%
\   \label{weinpot}
\end{equation}
There are three different $d_{ij}\ e^{i\theta _{ij}}$ terms, 
and only these can be
complex. In this case, the matrices $Y$, $\tilde{Z}$ and $\hat{Z}$ are
diagonal, hence the invariant written for the general 
case in Eq. (\ref{higgs3i}) is automatically zero. 
Yet, as it was pointed out by Weinberg,
there can be CP violation, since
in general one cannot rotate away simultaneously the three phases 
$\theta_{ij}$
through phase redefinitions of the three Higgs doublets. This implies that,
in this case, a CP-odd invariant will have to include terms 
of the form $Z_{1212}Z_{2323}Z_{3131}$ so as to be sensitive 
to the CP-violating phase
that cannot be rotated away. As a result, the simplest HBT 
invariant relevant to this
model must contain $Z_{ak}^{nq}Z_{nq}^{rs}Z_{rs}^{ka}$. However, such a
simple invariant is real. The simplest way out is to insert the two
diagonal matrices $Y$ and $\hat{Z}$ between $Z$'s. Thus, a relevant CP-odd
invariant is given by:
\begin{equation}
I_{2}^{W}=\mathrm{Im}[Z_{ak}^{nq}\ Y_{n}^{m}\ Z_{mq}^{rs}\ \widehat{Z}
_{r}^{t}\ Z_{ts}^{ka}] \ .  \label{weininv1}
\end{equation}
Its explicit form is:
\begin{equation}
\begin{array}{r}
I_{2}^{W}=d_{12}d_{13}d_{23}\ [(\ m_{3}-m_{2})(a_{11}-b_{23})-(\
m_{3}-m_{1})(a_{22}-b_{13})+ \\ 
\\ 
+(\ m_{2}-m_{1})(a_{33}-b_{12})]\sin (\theta _{12}-\theta _{13}+
\theta _{23}) \ .
\end{array}
\label{weinin2}
\end{equation}
Assuming non-degenerate values for the $m_i$, a non-vanishing $I_{2}^{W}$
indicates a non-vanishing $(\theta _{12}-\theta _{13}+\theta _{23})$.

\subsubsection{Three Higgs doublets with NFC and soft CP breaking}
Let us assume that the $Z_{2}\times Z_{2}\times Z_{2}$ symmetry,
which implements NFC in the Higgs potential, is softly broken
by quadratic terms in the potential. In this case, one can also have
soft CP breaking. In other words, CP can be violated even if all
quartic terms are real. Obviously, in this case the matrix $Y$ is 
no longer diagonal, in this basis, and the invariant $I_s$ 
defined in Eq.~(\ref{higgs3i}) does not
automatically vanish, being given by:
\begin{equation}
\begin{array}{r}
I_{s}=6(a_{22}-a_{33}+c_{12}-c_{13})(a_{33}-a_{11}+c_{23}-c_{12})
(a_{22}-a_{11}+
c_{23}-c_{13})\cdot 
\\ 
\\ 
\cdot \ y_{12}y_{13}y_{23}\ \sin [\varphi _{12}-
\varphi _{13}+\varphi _{23}] \ ,
\end{array}
\label{hig3inv1}
\end{equation}
where we have written the off-diagonal terms of the Hermitian matrix $Y$ as 
$Y_{ij}=y_{ij}\ e^{i\varphi _{ij}}$. 
On the other hand, in this model with soft CP breaking,
the CP-odd invariant $I_{2}^{W}$ 
vanishes. The reason for the vanishing of $I_{2}^{W}$ in this case is
the following: this invariant is also the imaginary part of the
trace of a multiplication of
the $Y$ matrix and a matrix made of combinations of $Z$'s. In the
Weinberg model this combination of $Z$'s is diagonal and, in addition,
it is real in the soft CP-breaking case. The matrix $Y$ is Hermitian,
hence the trace of the product of these two matrices is real.
This illustrates the usefulness of CP-odd HBT 
invariants in determining the character of CP breaking (hard or soft)
when the Higgs potential is written in an arbitrary basis, where
that character may not be transparent.

\section{Summary and conclusions}
We have shown how CP-odd HBT invariants can play an important r\^ ole
in the study of CP properties of Higgs systems. In order to derive
these invariants we first consider the most general CP
transformation, which leaves the Higgs kinetic-energy term invariant
and then derive the restrictions implied by CP invariance
on the quadratic and quartic scalar couplings. 
The method used in this paper to analyse CP invariance of the
Higgs potential does not require finding the actual CP transformation
under which the scalar potential is invariant. This renders this method
particularly useful, since such a transformation
may not be trivial and may become quite
complicated as the number of Higgs doublets increases.
The HBT invariants offer the advantage that they can be
directly evaluated in any Higgs basis. We have also illustrated
how the HBT invariants can be used to find out whether
there is soft or hard CP breaking in a given Higgs potential.
We have studied in detail only the case of two and three Higgs doublets,
but our approach can be readily extended to an arbitrary number
of these. One last comment is in order: throughout the paper,
we have only discussed the Higgs sector. It is clear that one can
readily extend the analysis to the remainder of the Lagrangian,
in particular the fermion sector. More specifically, one can
construct quantities involving both Yukawa and Higgs couplings
which are invariant under basis transformations comprising both 
HBT and fermion basis transformations.

\section*{Acknowledgements}

The authors thank the CERN Physics Department (PH) Theory (TH) for the warm
hospitality. This work was partially supported by CERN and by Funda\c c\~ao
para a Ci\^encia e a Tecnologia (FCT) (Portugal) through the Projects
PDCT/FP/FNU/ 50250/2003, PDCT/FP/FAT/50167/2003 and 
CFTP-FCT UNIT 777, which are partially funded through POCTI (FEDER).


\begin{thebibliography}{99}

\bibitem{Lee:1973iz} 
T.~D.~Lee, %``A Theory Of Spontaneous T Violation,''
Phys.\ Rev.\ D \textbf{8} (1973) 1226. %%CITATION = PHRVA,D8,1226;%%

\bibitem{Branco} 
G.~C.~Branco, 
%``Spontaneous CP Violation In Theories With More Than Four Quarks,''
Phys.\ Rev.\ Lett.\ \textbf{44} (1980) 504; %%CITATION = PRLTA,44,504;%%
G.~C.~Branco, 
%``Spontaneous CP Nonconservation And Natural Flavor Conservation: A Minimal
%Model,''
Phys.\ Rev.\ D \textbf{22} (1980) 2901; %%CITATION = PHRVA,D22,2901;%%
G.~C.~Branco, A.~J.~Buras and J.~M.~Gerard,
%``CP Violation In Models With Two And Three Scalar Doublets,''
Nucl.\ Phys.\ B {\bf 259} (1985) 306.
%%CITATION = NUPHA,B259,306;%%

\bibitem{colliders}
%\bibitem{Boz:2002yz}
M.~Boz and N.~K.~Pak,
%``Explicit CP violation in the general two-doublet model, with real CKM
%matrix,''
Phys.\ Rev.\ D {\bf 65} (2002) 075014;
%%CITATION = PHRVA,D65,075014;%%
%\bibitem{Iltan:2001vg}
E.~O.~Iltan,
%``Top quark electric and chromo electric dipole moments in the general  two
%Higgs doublet model,''
Phys.\ Rev.\ D {\bf 65} (2002) 073013
[arXiv:hep-ph/0111038];
%%CITATION = HEP-PH 0111038;%%
%\bibitem{Carena:2002bb}
M.~Carena, J.~R.~Ellis, S.~Mrenna, A.~Pilaftsis and C.~E.~M.~Wagner,
%``Collider probes of the MSSM Higgs sector with explicit CP violation,''
Nucl.\ Phys.\ B {\bf 659} (2003) 145
[arXiv:hep-ph/0211467];
%%CITATION = HEP-PH 0211467;%%
%\bibitem{Khater:2003wq}
W.~Khater and P.~Osland,
%``CP violation in top quark production at the LHC and two-Higgs-doublet
%models,''
Nucl.\ Phys.\ B {\bf 661} (2003) 209
[arXiv:hep-ph/0302004];
%%CITATION = HEP-PH 0302004;%%
%\bibitem{Diaz-Cruz:2004pj}
J.~L.~Diaz-Cruz, R.~Noriega-Papaqui and A.~Rosado,
%``Measuring the fermionic couplings of the Higgs boson at future colliders as
%a probe of a non-minimal flavor structure,''
arXiv:hep-ph/0410391;
%%CITATION = HEP-PH 0410391;%%
%\bibitem{Weiglein:2004hn}
G.~Weiglein {\it et al.}  [LHC/LC Study Group],
%``Physics interplay of the LHC and the ILC,''
arXiv:hep-ph/0410364, and references therein;
%%CITATION = HEP-PH 0410364;%%
%\bibitem{Group:2004sz}
C.~P.~W.~Group {\it et al.},
%``Physics at the CLIC multi-TeV linear collider,''
arXiv:hep-ph/0412251, and references therein;
%%CITATION = HEP-PH 0412251;%%
A.~Sopczak,
%``Higgs Physics: from LEP to a Future Linear Collider,''
arXiv:hep-ph/0502002.
%%CITATION = HEP-PH 0502002;%%



\bibitem{Bernabeu:1986fc} 
J.~Bernabeu, G.~C.~Branco and M.~Gronau, 
%``CP Restrictions On Quark Mass Matrices,''
Phys.\ Lett.\ B \textbf{169} (1986) 243. %%CITATION = PHLTA,B169,243;%%

\bibitem{many1} 
G.~C.~Branco and M.~N.~Rebelo, 
%``Weak Basis Invariant Conditions For CP Conservation In SU(2)-L X SU(2)-R X
%U(1)-(B-L) Models,''
Phys.\ Lett.\ B \textbf{173} (1986) 313; %%CITATION = PHLTA,B173,313;%%
M.~Gronau, A.~Kfir and R.~Loewy, 
%``Basis Independent Tests Of CP Violation In Fermion Mass Matrices,''
Phys.\ Rev.\ Lett.\ \textbf{56} (1986) 1538; %%CITATION = PRLTA,56,1538;%%
G.~C.~Branco and V.~A.~Kostelecky, 
%``CP Violation In Supergravity Models,''
Phys.\ Rev.\ D \textbf{39} (1989) 2075; %%CITATION = PHRVA,D39,2075;%%
G.~C.~Branco, M.~N.~Rebelo and J.~W.~F.~Valle, 
%``Leptonic CP Violation With Massless Neutrinos,''
Phys.\ Lett.\ B \textbf{225} (1989) 385; %%CITATION = PHLTA,B225,385;%
F.~del Aguila and J.~A.~Aguilar-Saavedra, 
%``Invariant formulation of CP violation for four quark families,''
Phys.\ Lett.\ B \textbf{386} (1996) 241 [arXiv:hep-ph/9605418]; 
%%CITATION = HEP-PH 9605418;%%
F.~del Aguila, J.~A.~Aguilar-Saavedra and M.~Zralek, 
%``Invariant analysis of CP violation,''
Comput.\ Phys.\ Commun.\ \textbf{100} (1997) 231 [arXiv:hep-ph/9607311]; 
%%CITATION = HEP-PH 9607311;%%
J.~A.~Aguilar-Saavedra, 
%``Measure of the size of CP violation in extended models,''
J.\ Phys.\ G \textbf{24} (1998) L31 [arXiv:hep-ph/9703461]; 
%%CITATION = HEP-PH 9703461;%%
A.~Pilaftsis, 
%``CP violation and baryogenesis due to heavy Majorana neutrinos,''
Phys.\ Rev.\ D \textbf{56} (1997) 5431 [arXiv:hep-ph/9707235]; 
%%CITATION = HEP-PH 9707235;%%
O.~Lebedev, %``CP-violating invariants in supersymmetry,''
Phys.\ Rev.\ D \textbf{67} (2003) 015013 [arXiv:hep-ph/0209023]; 
%%CITATION = HEP-PH 0209023;%%
S.~Davidson and R.~Kitano, %``Leptogenesis and a Jarlskog invariant,''
JHEP \textbf{0403} (2004) 020 [arXiv:hep-ph/0312007]; 
%%CITATION = HEP-PH 0312007;%%
F.~J.~Botella, M.~Nebot and O.~Vives, 
%``Invariant approach to flavour-dependent CP-violating phases in the MSSM,''
arXiv:hep-ph/0407349. %%CITATION = HEP-PH 0407349;%%

\bibitem{Branco:1999fs} 
G.~C.~Branco, L.~Lavoura and J.~P.~Silva, ``CP
violation,'' International Series of Monographs on Physics, No. 103, Oxford
University Press. Oxford, UK: Clarendon (1999), 511 pp.

\bibitem{higgs} 
L.~Lavoura and J.~P.~Silva, 
%``Fundamental CP violating quantities in a SU(2) x U(1) model with many Higgs
%doublets,''
Phys.\ Rev.\ D \textbf{50} (1994) 4619 [arXiv:hep-ph/9404276]; 
%%CITATION = HEP-PH 9404276;%%
F.~J.~Botella and J.~P.~Silva, 
%``Jarlskog - like invariants for theories with scalars and fermions,''
Phys.\ Rev.\ D \textbf{51} (1995) 3870 [arXiv:hep-ph/9411288].
%%CITATION = HEP-PH 9411288;%%

\bibitem{cern} 
I.~F.~Ginzburg and M.~Krawczyk,
%``Symmetries of two Higgs doublet model and CP violation,''
arXiv:hep-ph/0408011.
%%CITATION = HEP-PH 0408011;%%
J. Gunion and H. Haber in: Talk by J. Gunion at the Workshop
``CP studies and non-standard Higgs physics'', CERN, 2-3 December 2004,
http://kraml.home.cern.ch/kraml/cpnsh/

\bibitem{Branco:1985aq} 
G.~C.~Branco and M.~N.~Rebelo, 
%``The Higgs Mass In A Model With Two Scalar Doublets And Spontaneous CP
%Violation,''
Phys.\ Lett.\ B \textbf{160} (1985) 117. %%CITATION = PHLTA,B160,117;%%

\bibitem{Glashow:1976nt} 
S.~L.~Glashow and S.~Weinberg, 
%``Natural Conservation Laws For Neutral Currents,''
Phys.\ Rev.\ D \textbf{15} (1977) 1958. %%CITATION = PHRVA,D15,1958;%%


\bibitem{Weinberg:1976hu} 
S.~Weinberg, %``Gauge Theory Of CP Violation,''
Phys.\ Rev.\ Lett.\ \textbf{37} (1976) 657. %%CITATION = PRLTA,37,657;%%





\end{thebibliography}
\end{document}